\documentclass[preprint,aps]{revtex4-1}

\usepackage{graphicx}

\begin{document}

\title{Unusual Electronic Structure of Dirac Material BaMnSb$_2$ Revealed by Angle-Resolved Photoemission Spectroscopy}

\author{Hongtao Rong$^{1,2,\sharp}$, Liqin Zhou$^{1,2,\sharp}$, Junbao He$^{1,2,3,\sharp}$, Chunyao Song$^{1,2,\sharp}$, \\Yu Xu$^{1,2}$, Yongqing Cai$^{1,2}$, Cong Li$^{1,2}$, Qingyan Wang$^{1,2}$, Lin Zhao$^{1,2,4}$,\\ Guodong Liu$^{1,2,4}$, Zuyan Xu$^{5}$, Genfu Chen$^{1,2}$, Hongming Weng$^{1,2}$ and X. J. Zhou$^{1,2,4,6,*}$}

\affiliation{
\\$^{1}$National Lab for Superconductivity, Beijing National Laboratory for Condensed Matter Physics, Institute of Physics, Chinese Academy of Sciences, Beijing 100190, China
\\$^{2}$University of Chinese Academy of Sciences, Beijing 100049, China
\\$^{3}$Henan International Joint Laboratory of MXene Materials Microstructure, College of Physics and Electronic Engineering, Nanyang Normal University, Nanyang 473061, China
\\$^{4}$Songshan Lake Materials Laboratory, Dongguan 523808, China
\\$^{5}$Technical Institute of Physics and Chemistry, Chinese Academy of Sciences, Beijing 100190, China
\\$^{6}$Beijing Academy of Quantum Information Sciences, Beijing 100193, China
\\$^{*}$Corresponding author: XJZhou@iphy.ac.cn.
}
\date{\today}

\begin{abstract}
High resolution angle resolved photoemission measurements and band structure calculations are carried out to study the electronic structure of BaMnSb$_2$. All the observed bands are nearly linear that extend to a wide energy range. The measured Fermi surface mainly consists of one hole pocket around $\Gamma$ and a strong spot at Y which are formed from the crossing points of the linear bands. The measured electronic structure of BaMnSb$_2$ is unusual and deviates strongly from the band structure calculations. These results will stimulate further efforts to theoretically understand the electronic structure of BaMnSb$_2$ and search for novel properties in this Dirac material.
\end{abstract}


\maketitle

Topological materials, including topological insulators, topological semimetals and topological superconductors, have attracted much attention due to their unique electronic structure, exotic physical properties and potential applications\cite{MZHasan,XLQiTIS,ABansil,CFangTNL,BHYan,NPArmitage,YTokura}. Magnetic topological materials, with the introduction of magnetism to break the time-reversal symmetry, can generate novel topological phases such as the quantum anomalous Hall state\cite{XLQi,XGWan,RYu,CZChang,MMOtrokov,YGong,DQZhang,YFXu}. The AMnPn$_2$ (A = Ca, Sr, Ba, Eu or Yb; Pn= Bi or Sb) family has provided a desirable platform to search for intrinsic magnetic topological materials\cite{YFeng,SBorisenkoYMB,RKealhoferYMS,JParkSMB,HRyuBMB,HSakaiBMS,JYLiuBMS,SVRamankuttySMS,JRSohEMS,JYLSMS,HMasudaEMB}. In these materials, the Bi or Sb layer can host Dirac dispersion and the Mn sublattice can provide magnetic environment. The introduction of Eu or Yb on the A site can further manipulate the magnetic structure in the system. In addition to the magnetic structure, it has been found that the electronic structure and topological properties also depend sensitively on the crystal structure\cite{HTRongCMS}. Among all the compounds in the AMnPn$_2$, BaMnSb$_2$ is unique because it has a crystal structure that is distinct from other members in the family. While there have been some investigations on BaMnSb$_2$\cite{JYLiuBMS0,SLHuangBMS,HSakaiBMS,JBYuBMS,JYLiuBMS}, the detailed study of its electronic structure is still limited\cite{HSakaiBMS,JYLiuBMS}.

In this paper, we have carried out detailed high resolution angle resolved photoemission spectroscopy  (ARPES) measurements and band structure calculations to study the electronic structure of BaMnSb$_2$. The observed Fermi surface mainly consists of one hole pocket around $\Gamma$ and a strong spot at Y. The measured band structures are dominated by nearly linear bands that extend to a wide energy range. The measured electronic structures are unusual that deviate obviously from the band structure calculations.

High-quality single crystals of BaMnSb$_{2}$ were grown by flux method\cite{JBHeCMS}. The elements Ba, Mn, and Sb, were mixed in the ratio of Ba: Mn: Sb = 1: 1: 4, put into an alumina crucible, and sealed in a quartz tube. The quartz tube was heated slowly to 900 $^{\circ}C$, held for 20 hours, and then cooled to 700 $^{\circ}C$ at a rate of 1 $^{\circ}C$/hour, where the excess flux was removed using a centrifuge. Shiny platelike single crystals with a typical dimension of $5 \times 5 \times 1$ mm$^{3}$ were obtained. We carried out single crystal x-ray diffraction to analyse the crystal structure of the samples at 273 K. It crystalizes in an orthorhombic structure (space group: Imm2) with lattice constants a=24.258 {\AA}, b=4.4827 {\AA} and c=4.506 {\AA}, consistent with the previous reports\cite{HSakaiBMS,JYLiuBMS}. ARPES measurements were performed at our lab-ARPES system equipped with 21.218 eV Helium discharge lamp and Scienta DA30L electron energy analyzer\cite{GDLiu,XJZhou}. The energy resolution was set at $\sim$20 meV. The angular resolution was $\sim0.3^{\circ}$. The Fermi level is referenced by measuring on the Fermi edge of a clean polycrystalline gold that is electrically connected to the sample. All the samples were cleaved {\it in situ} and measured in ultrahigh vacuum with a base pressure better than $5\times10^{-11}$ mbar.

The first-principle calculations to simulate the electronic structures of BaMnSb$_{2}$ were implemented by using the Vienna {\it ab initio} simulation package (VASP)\cite{GKresse}. The generalized gradient approximation (GGA) in the Perdew-Burke-Ernzerhof (PBE) type was selected to describe the exchange-correlation function\cite{JPPerdew}. The Brillouin zone (BZ) integration was sampled by $3 \times 11 \times 11$ k mesh and the cut-off energy was set to 500 eV. Spin-orbital coupling (SOC) was taken into account. The tight-binding model of BaMnSb$_{2}$ was constructed by using the Wannier90 with Ba 5$d$ orbitals, Sb 5$p$ orbitals and Mn 3$d$ orbitals, which is based on the maximally-localized Wannier functions (MLWF)\cite{AAMostofi}. The bulk Fermi surfaces of BaMnSb$_{2}$ were calculated by the WannierTools package\cite{QSWu}.

The crystal structure of BaMnSb$_{2}$ is shown in Fig. 1a. It consists of an alternate stacking of MnSb$_{4}$ layer and Ba-Sb-Ba layer along the {\it a} axis. In the central Ba-Sb-Ba layer, the Sb atoms in the Sb sheet form zig-zag chains with the chain direction along the {\it b} axis (Fig. 1b). The Ba atoms above and below the Sb sheet are coincident. Fig. 1c shows the measured Fermi surface mapping of BaMnSb$_{2}$ at 30 K. The main features observed consist of a circular Fermi surface sheet around $\Gamma$ and a strong spot at Y. No features are observed at X. Fig. 1d shows the constant energy contours at different binding energies. At low binding energies below $\sim$100 meV, one circle can be observed around $\Gamma$. When the binding energy is higher than $\sim$200 meV, combined with the detailed band structure analysis near $\Gamma$(Fig. 2) and near Y(Fig. 3), three main features appear around $\Gamma$: an outer sheet $\alpha$, an inner vertical ellipse $\beta$ and another inner horizontal ellipse $\gamma$. With increasing binding energy, the outer $\alpha$ sheet increases in its area significantly while the two inner ellipses $\beta$ and $\gamma$ get elongated along their long axis. In the mean time, as seen in Fig. 1d, the feature at Y stays as a strong spot up to the binding energy of $\sim$200 meV and it disintegrates at higher binding energies.

In Fig. 2, we zoom in onto the detailed electronic structure around $\Gamma$ point in BaMnSb$_{2}$. Fig. 2(a-d) shows the band structures measured along different momentum cuts around $\Gamma$. In all these measurements, mainly two sets of bands are observed, labeled as $\alpha$ and $\beta$ bands in Fig. 2b for the cut 6 along the $\Gamma$-X direction and labeled as $\beta$ and $\gamma$ bands in Fig. 2d for the cut 12 along the $\Gamma$-Y direction. These $\alpha$, $\beta$ and $\gamma$ bands are nearly linear in dispersion that extends to a wide energy range up to 0.9 eV (Fig. 2b). As seen in Fig. 2(a-b) for the momentum cuts parallel to $\Gamma$-X, the $\beta$ band is very steep; the corresponding Fermi velocity is $\sim$13 eV$\cdot${\AA} for the $\beta$ band in Fig. 2b. The two linear bands $\alpha$ and $\beta$ intersect, as shown in Fig. 2a and 2b, and the energy position of the crossing point varies for different momentum cuts. For the momentum cuts 4 to 8, we find that the crossing point is closed the Fermi level. These Dirac cone-like bands give rise to the formation of the hole-like Fermi surface around $\Gamma$ as shown in Fig. 2e. When the momentum cuts move away further from $\Gamma$ (cuts 1-3 and cuts 9-11), the crossing point of the $\alpha$ and $\beta$ bands shifts down from the Fermi level and the spectral weight at the Fermi level gets depleted. The $\alpha$ band and $\beta$ band observed in Fig. 2 contribute to the formation of the $\alpha$ sheet and $\beta$ sheet in the constant energy contours at high binding energies above 200\.meV in Fig. 1d. The band structure for the cut 12 along the $\Gamma$-Y direction, shown in Fig. 2d, also consists of two main bands around $\Gamma$. In this case, combined with the constant energy contours in Fig. 1d, the outer band can be assigned to $\beta$ that forms the $\beta$ sheet in Fig. 1d while the inner $\gamma$ band contributes to the formation of the $\gamma$ sheet in Fig. 1d. In addition to the main linear bands $\alpha$ and $\beta$, there is another band observed in the measured band structure for the momentum cut 6 along the $\Gamma$-X direction (Fig. 2c). This band is flat; it does not show up in the MDC second derivative image in Fig. 2b but becomes clear in the EDC second derivative image in Fig. 2c. The band lies about 130\.meV below the Fermi level. The combined MDC and EDC second derivative analyses provide a complete picture of the band structure around $\Gamma$ (Fig. 2a, 2b and 2c).

Figure 3 shows the zoomed-in constant energy contours and the detailed electronic structures of BaMnSb$_{2}$ around the Y point. In all the bands measured along different momentum cuts around Y in Fig. 3d and 3e, we observed mainly one set of bands, named as $\delta$ in Fig. 3e. These $\delta$ bands are nearly linear over a large energy range ($\sim$0.9 eV in Fig. 3e). For the momentum cut 6 along the Y-M direction, the two linear bands have a Fermi velocity of $\sim$2.9 eV$\cdot${\AA} and cross each other near the Fermi level (Fig. 3e). When the momentum cuts move away from cut 6 (cut 5- cut 1 and cut 7- cut 11 in Fig. 3d), the spectral weight gets suppressed more and more below the Fermi level. Such a momentum-dependent band evolution produces a strong spot at Y in the measured Fermi surface (Fig. 3a) and the evolution of the constant energy contours around Y as shown in Fig. 1d, Fig. 3b and 3c. In addition to the main linear bands, there is another band observed in the measured band structure for the momentum cut 6 along the Y-M direction (Fig. 3f). This band is hole-like with its top at $\sim$300 meV. It does not show up in the MDC second derivative image in Fig. 3e but becomes visible in the EDC second derivative image in Fig. 3f.

In order to understand our measured results, we carried out detailed band structure calculations of BaMnSb$_2$ as shown in Fig. 4. In the calculations, we took a G-type magnetic structure in BaMnSb$_2$ (Fig. 4a) where both the interlayer and intralayer couplings between two nearest moments in the MnSb$_4$ layer are antiferromagnetic\cite{JYLiuBMS0}. Fig. 4c shows the overall calculated band structure of BaMnSb$_2$ without considering the spin-orbital coupling (SOC). The low energy band structure near the $\Gamma$ point mainly consists of two M-shaped bands. The band structure near the Y point is dominated by a cusp structure below the Fermi level and a dip structure above the Fermi level. The position of the cusp and dip structures deviates from the Y point along the Y-M direction. Even without considering SOC, there is a prominent gap between the dip and the cusp that is about 380 meV. When SOC is taken into account in the band structure calculations, overall it causes the lifting of the band degeneracy and band splitting as seen in Fig. 4d and 4e. The effect of SOC on the low energy band structure near the $\Gamma$ point is small but dramatic changes of the band structure occur near Y point (Fig. 4d and 4e). The initial one band of the cusp and dip structures in Fig. 4c splits into multiple bands. In the meantime, both structures move towards the Fermi level and develop into a Dirac cone-like band with a much reduced gap of $\sim$130 meV. The low energy electronic structure of BaMnSb$_{2}$ is rather two-dimensional. When $k_{z}$ moves from 0 (Fig. 4d) to $\pi/a$ (Fig. 4e), the bands near the Y point are barely changed, although there is a slight change in the bands near the $\Gamma$ point.

Now we compare the calculated and measured electronic structures of BaMnSb$_{2}$. Fig. 5a and 5b compare the calculated (Fig. 5a) and measured (Fig. 5b) band structures along several high symmetry directions. Fig. 5c-5f compares the calculated (Fig. 5c and 5d) and measured (Fig. 5e and 5f) Fermi surface and constant energy contour. Considering the observation of the feature at Y, an experimental Fermi level E$_{F(exp)}$ is taken which is 0.16 eV below the Fermi level in the calculation as show in Fig. 5a. The measured electronic structure of BaMnSb$_{2}$ shows some consistence with the calculations on a few main aspects. The first is the hole like bands around $\Gamma$ and Dirac-like band around Y (Fig. 5a and 5b). Second, the Fermi surface consists of the hole like sheet around $\Gamma$ and some peculiar features around Y (Fig. 5c and 5e). Third, the observed features in the measured constant energy contour (Fig. 5f) show some resemblance to those in the calculated constant energy contour (Fig. 5d). However, the overall calculated electronic structure shows some strong deviations from the measured results. First, the number of the observed bands is much less than that of the calculated bands both around the $\Gamma$ and Y points. Second, the observed bands are nearly linear that extend to a large energy range of $\sim$0.9 eV (Fig. 5b). It is hard to find the corresponding bands in the calculations (Fig. 5a). Third, although there are some similarities in the calculated and measured Fermi surface and constant energy contours, there remain significant differences on the size of the Fermi surface and the features in the constant energy contour. In some other compounds like SrMnBi$_2$, CaMnBi$_2$ and CaMnSb$_2$, the measured electronic structure is found to be in an excellent agreement with the band structure calculations\cite{YFeng,HTRongCMS}. The measured electronic structure of BaMnSb$_2$ is unusual that it can not be described properly by the band structure calculations.

In summary, we have carried out detailed high resolution ARPES measurements and band structure calculations to study the electronic structure of BaMnSb$_2$. The measured electronic structure of BaMnSb$_2$ is unusual in three aspects. First, all the measured bands are nearly linear that extend to a wide energy range. Second, the observed Fermi surface mainly consists of one hole pocket around $\Gamma$ and a strong spot at Y. The Fermi surface is formed from the crossing points of the linear bands. Third, the measured electronic structure deviates strongly from the band structure calculations. These results will stimulate further efforts to theoretically understand the electronic structure of BaMnSb$_2$ and search for novel properties in this Dirac material.

\vspace{3mm}

\textbf{Acknowledgments}
This work is supported by the National Key Research and Development Program of China (Nos. 2016YFA0300600, 2018YFA0305602, 2016YFA0300300 and 2017YFA0302900), the National Natural Science Foundation of China (Nos. 11974404, 11888101, 11922414 and 11404175), the Strategic Priority Research Program (B) of the Chinese Academy of Sciences (Nos. XDB33000000 and  XDB25000000), the Youth Innovation Promotion Association of CAS (No. 2017013)and the Natural Science Foundation of Henan Province (Nos. 182300410274 and 202300410296). The theoretical calculations is supported by the National Natural Science Foundation of China (Grant Nos. 11674369, 11865019 and 11925408), the Beijing Natural Science Foundation (Grant No. Z180008), Beijing Municipal Science and Technology Commission (Grant No. Z191100007219013), the National Key Research and Development Program of China (Grant Nos. 2016YFA0300600 and 2018YFA0305700), the K. C. Wong Education Foundation (Grant No. GJTD-2018-01) and the Strategic Priority Research Program of Chinese Academy of Sciences (Grant No. XDB33000000)

\vspace{3mm}

\textbf{Author contributions}
  H.T.R, L.Q.Z, J.B.H and C.Y.S contribute equally to this work. X.J.Z. and H.T.R. proposed and designed the research. H.T.R. and Y.X. carried out the ARPES experiments. L.Q.Z. and H.M.W. contributed in the band structure calculations. J.B.H. and G.F.C. prepared the samples; H.T.R., C.Y.S., Y.X., Y.Q.C., H.C., C.L., Q.Y.W., L.Z., G.D.L. Z.Y.X. and X.J.Z. contributed to the development and maintenance of Laser-ARPES systems, the data analysis and the related software development. X.J.Z. and H.T.R. wrote the paper. All authors participated in discussions and comments on the paper.

\vspace{3mm}
\textbf{Additional information}

\vspace{3mm}
\textbf{Competing financial interests}:
 The authors declare no competing financial interests.

\newpage

\begin{figure*}[t]
\begin{center}
\includegraphics[width=0.99\columnwidth,angle=0]{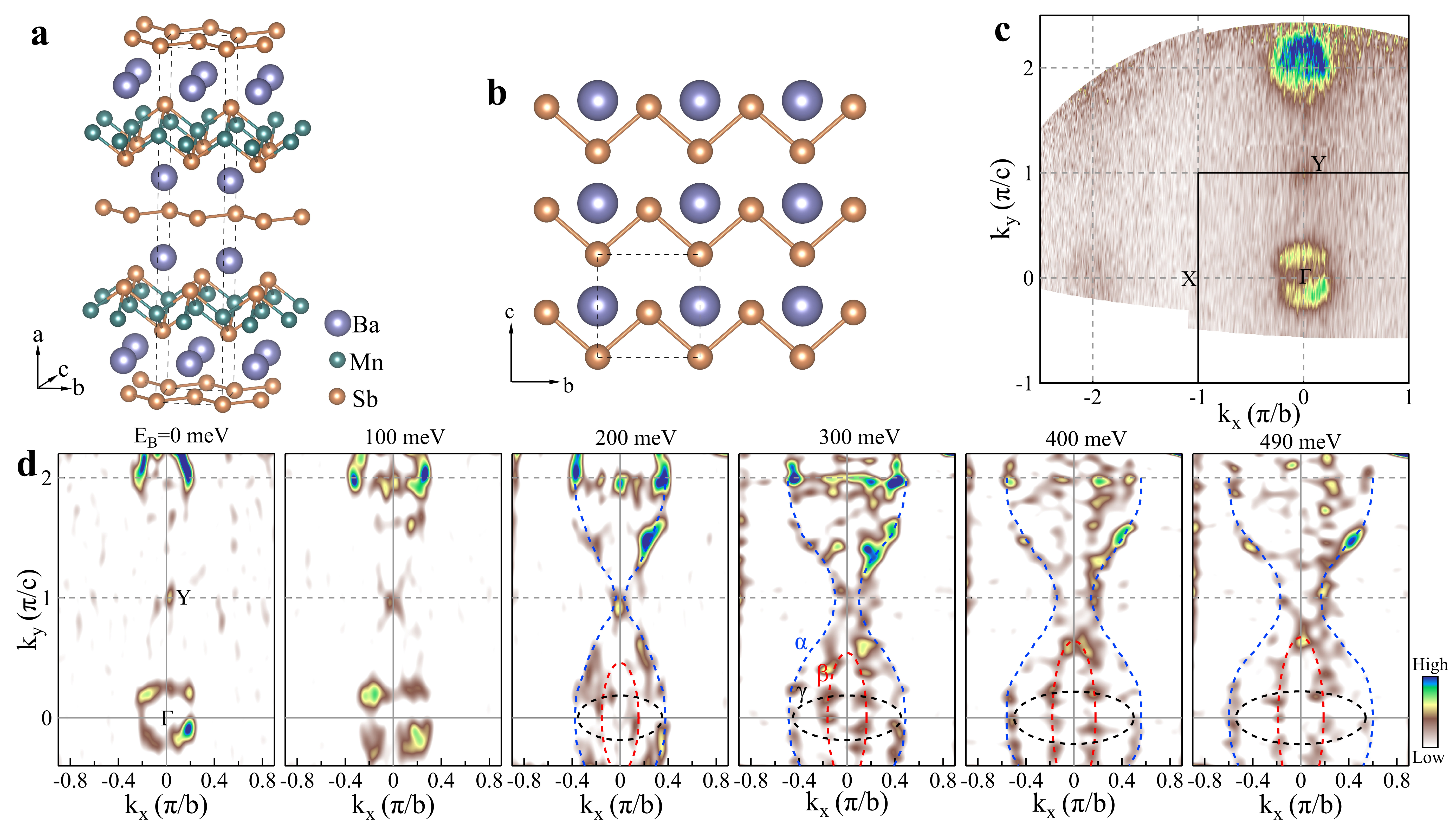}
\end{center}
\begin{center}
\caption{{\bf Fermi surface mapping and constant energy contours of BaMnSb$_{2}$.} (a) Crystal structure of BaMnSb$_{2}$. The unit cell is indicated by the black dashed lines. (b) Top view of the central Ba-Sb-Ba layer which consists of the zig-zag chainlike structure along the b axis in the Sb layer. (c) Fermi surface mapping obtained by integrating the photoemission spectral weight over a [-20,20] meV energy window with respect to the Fermi level. (d) Constant energy contours at different binding energies of 0 meV, 100 meV, 200 meV, 300 meV, 400 meV and 490 meV. These are second derivative images obtained by integrating the spectral weight over a [-20,20] meV energy window with respect to each binding energy.
}
\end{center}
\end{figure*}

\begin{figure*}[t]
\begin{center}
\includegraphics[width=0.99\columnwidth,angle=0]{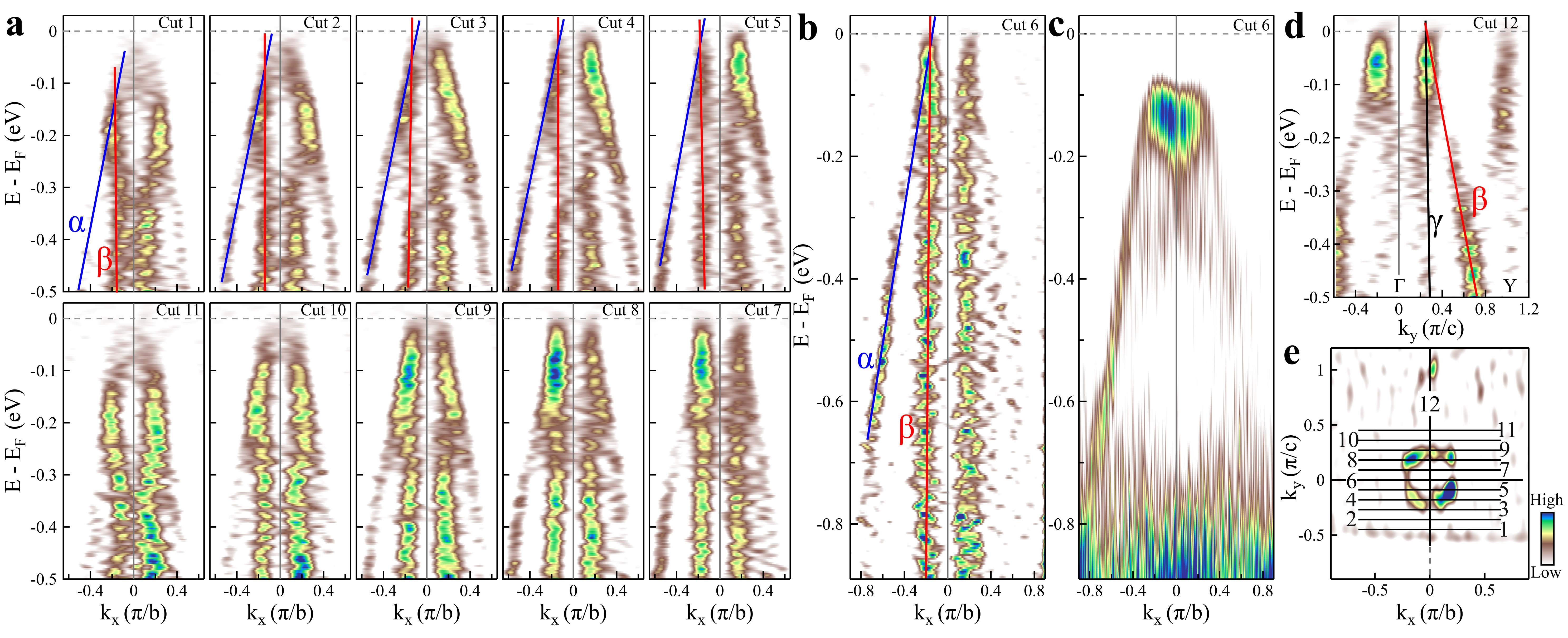}
\end{center}
\begin{center}
\caption{{\bf Hole-like Fermi surface and detailed band structures of BaMnSb$_2$ around $\Gamma$ measured at 30 K.} (a-b) Band structure measured along the $\Gamma$-X direction for various momentum cuts at different k$_{y}$s. These are the second derivative images with respect to momentum. The location of the momentum cuts, cut 1 to cut 11, is marked by black lines in (e). (c) Band structure for momentum cut 6. It is the second derivative image with respect to energy. (d) Band structure for momentum cut 12; it location is marked by black line in (e). It is the second derivative image with respect to momentum. (e) Fermi surface mapping around $\Gamma$. It is the second derivative image obtained by integrating the spectral weight over a [-20,20] meV energy window with respect to the Fermi level.
}
\end{center}
\end{figure*}

\begin{figure*}[t]
\begin{center}
\includegraphics[width=0.99\columnwidth,angle=0]{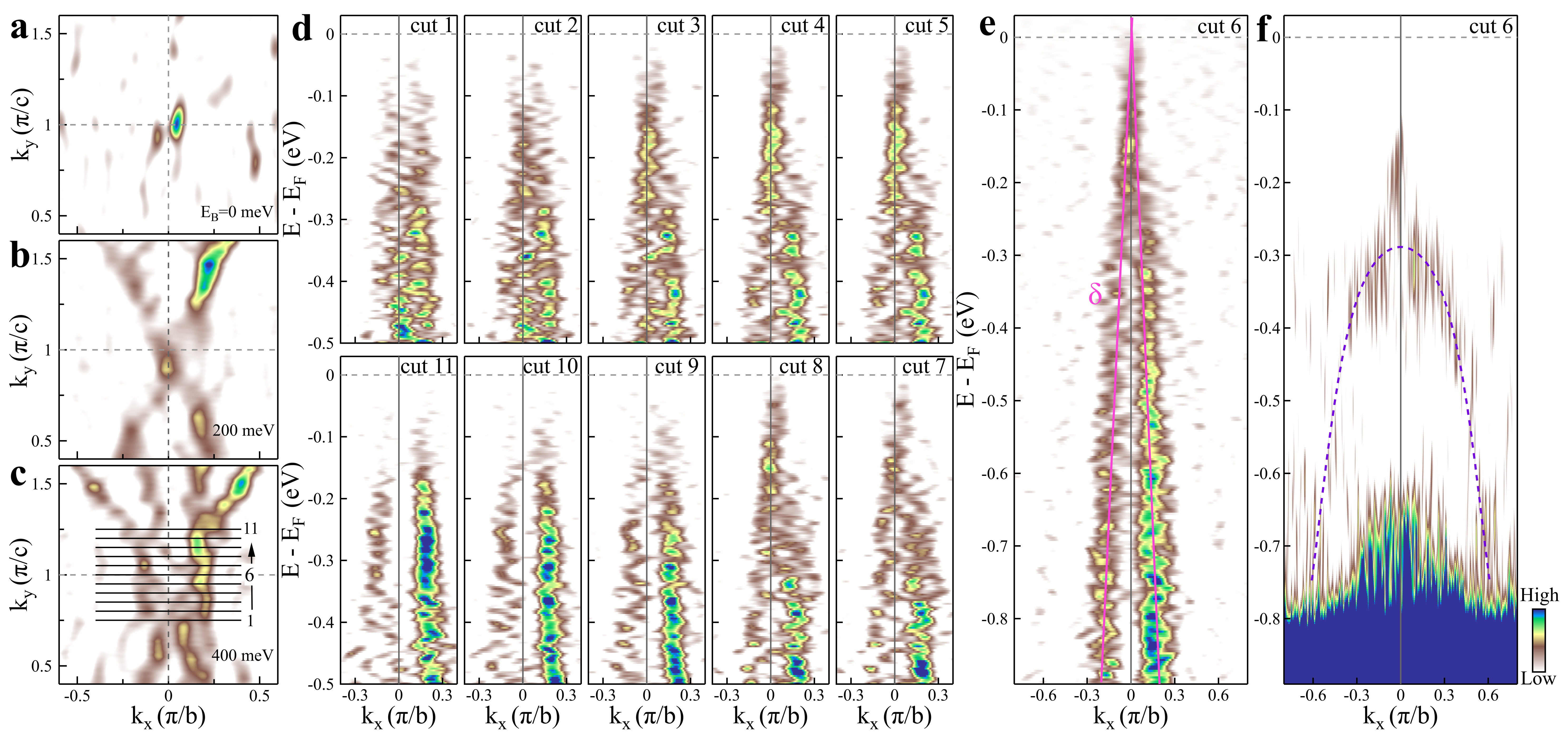}
\end{center}
\begin{center}
\caption{{\bf Fermi surface mapping and detailed band structures of BaMnSb$_2$ around Y measured at 30 K.} (a-c) Constant energy contours at binding energies of 0 meV (a), 200 meV (b) and 400 meV (c). These are second derivative images obtained by integrating the spectral weight over a [-20,20] meV energy window at binding energies of 0 meV, 200 meV and 400 meV. (d-e) Band structure measured along the $\Gamma$-X direction for various momentum cuts at different k$_{y}$s. These are the second derivative images with respect to momentum. The location of the momentum cuts, cut 1 to cut 11, is marked by black lines in (a). (f) Band structure for momentum cut 6. It is the second derivative image with respect to energy.
}
\end{center}
\end{figure*}

\begin{figure*}[t]
\begin{center}
\includegraphics[width=0.99\columnwidth,angle=0]{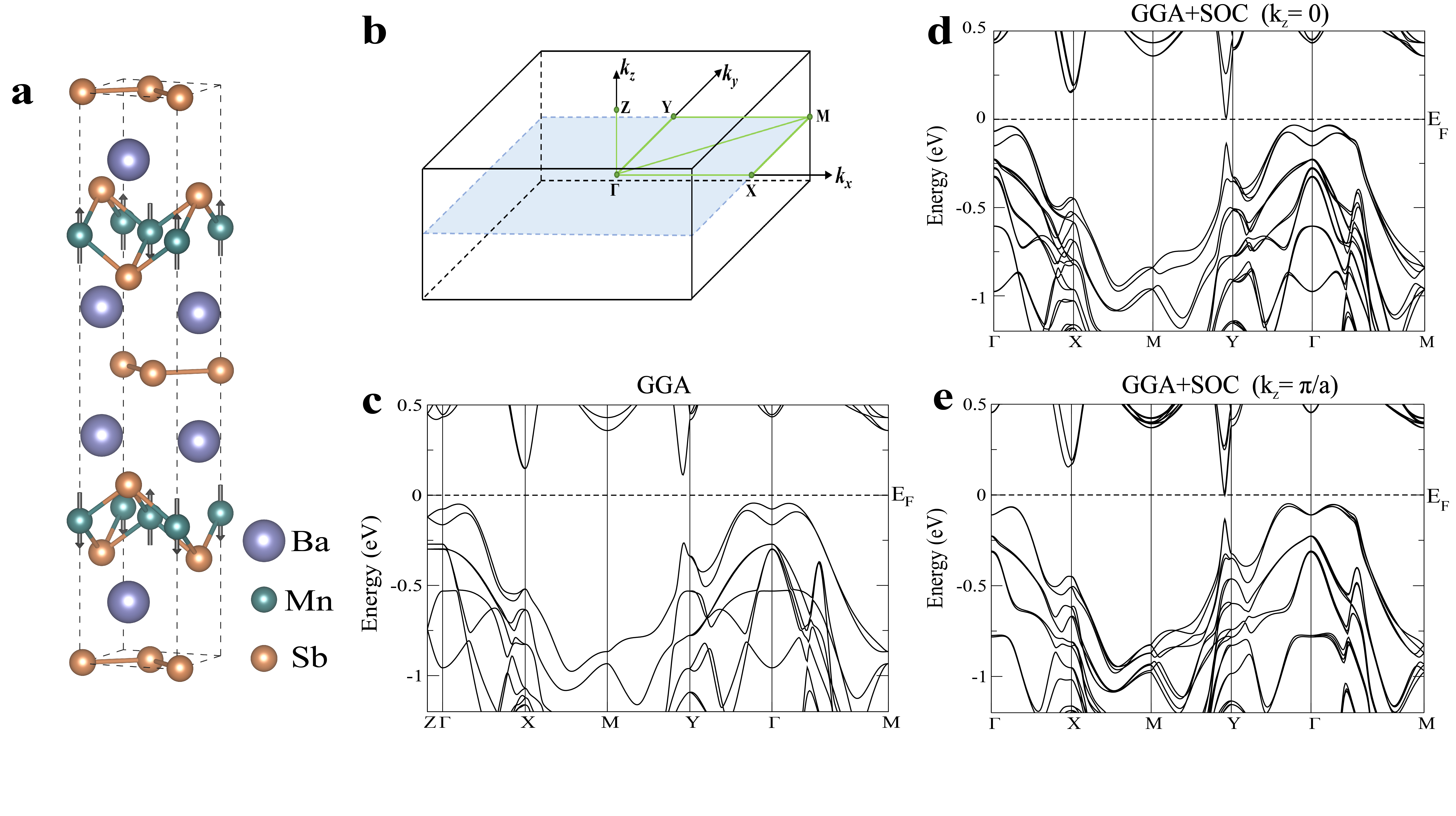}
\end{center}
\begin{center}
\caption{{\bf Calculated band structures of BaMnSb$_2$.} (a) Magnetic structure of BaMnSb$_2$. (b) Three dimensional Brillouin zone of BaMnSb$_2$. Green lines indicate high symmetry directions. (c,d) Calculated band structure of BaMnSb$_2$ without (c) and with (d) spin-orbit coupling at $k_{z}=0$. (e) Calculated band structure with spin-orbit coupling at $k_{z}=\pi/a$.
}
\end{center}
\end{figure*}

\begin{figure*}[t]
\begin{center}
\includegraphics[width=0.99\columnwidth,angle=0]{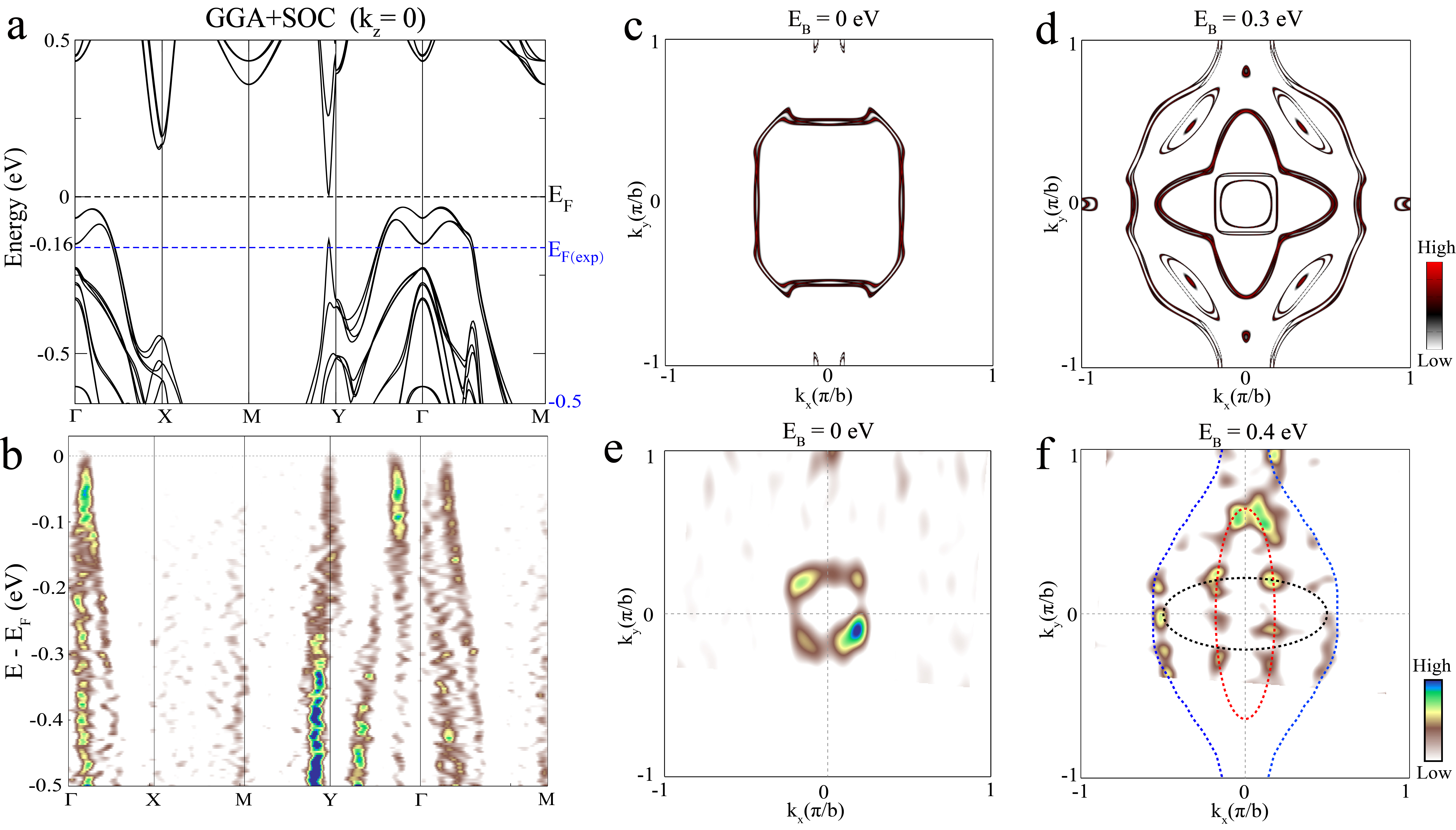}
\end{center}
\begin{center}
\caption{{\bf Comparison between the calculated and measured electronic structures of BaMnSb$_{2}$.} (a) Calculated band structure of BaMnSb$_2$ with spin-orbit coupling at $k_{z}=0$. (b) Measured band structure of BaMnSb$_2$. This is the second derivative image with respect to the momentum. (c-d) Calculated Fermi surface (c) and constant energy contour at a binding energy of 0.3 eV (d). (e-f) Measured Fermi surface (e) and constant energy contour at a binding energy of 0.4 eV (f). These are second derivative images obtained by integrating the spectral weight over a [-20,20] meV energy window with respect to each binding energy.
}
\end{center}
\end{figure*}

\end{document}